\documentclass[preprint,review,12pt]{elsarticle}

\usepackage{graphicx}
\usepackage{caption}

\usepackage{amssymb}

\usepackage{amsmath}

\usepackage[lined,ruled]{algorithm2e}

\usepackage{booktabs}
\usepackage{adjustbox}

\newcounter{bla}

\journal{Computer Physics Communications}

\begin{document}

\begin{frontmatter}



\title{A fast algorithm for computing a matrix transform used to detect trends in noisy data}


\author[a]{Dan Kestner}
\author[b]{Glenn Ierley}
\author[a]{Alex Kostinski\corref{author}}

\cortext[author] {Corresponding author.\\\textit{E-mail address:} djkestne@mtu.edu}
\address[a]{Physics Department, Michigan Technological University, 1400 Townsend Dr. Houghton MI. 49931}
\address[b]{Mathematics Department, Michigan Technological University, 1400 Townsend Dr. Houghton MI. 49931}

\begin{abstract}
A recently discovered universal rank-based matrix method to extract trends from noisy time series is described in [1] but the formula for the output matrix elements, implemented there as an open-access supplement MATLAB computer code, is ${\cal O}(N^4)$, with $N$ the matrix dimension. This can become prohibitively large for time series with hundreds of sample points or more. Based on recurrence relations, here we derive a much faster ${\cal O}(N^2)$ algorithm and provide code implementations in MATLAB and in open-source JULIA. In some cases one has the output matrix and needs to solve an inverse problem to obtain the input matrix. A fast algorithm and code for this companion problem, also based on the above recurrence relations, are given. Finally, in the narrower, but common, domains of (i) trend detection and (ii) parameter estimation of a linear trend, users require, not the individual matrix elements, but simply their accumulated mean value. For this latter case we provide a yet faster ${\cal O}(N)$ heuristic approximation that relies on a series of rank one matrices. These algorithms are illustrated on a time series of high energy cosmic rays with $N > 4 \times 10^4$.
\end{abstract}


\end{frontmatter}


{\bf PROGRAM SUMMARY}

\begin{small}
\noindent
{\em Program Title:} Pfromdata, QofP, mbasisandcoeffs, nonzerop, Qavgapprox, PofQ, Testing            \\
{\em Program Files doi:} http://dx.doi.org/xx.xxxxx/xxxxx.x (to be assigned by journal)        \\
{\em Licensing provisions:} MIT (Julia)         \\
{\em Programming language:} MATLAB and Julia        \\
{\em Nature of problem:} An order-rank data matrix and its transform to a stable form are used repeatedly to detect and/or extract trends from noisy data. An efficient yet accurate calculation of the matrix transform is therefore required.\\
{\em Solution method:} We introduce and apply an analytic recursion relation, which speeds up the execution of the matrix transform from ${\cal O}(N^4)$ arithmetic operations to ${\cal O}(N^2)$. Since this matrix transform is called often during optimization, our improvement allows for far shorter optimization times, for a given sample size. For example, a transform whose time is extrapolated to an unrealistic 75 days on a Dell personal laptop computer with a 2.2 GHz quad-core AMD processor running 32 bit MATLAB version R2015b on 64 bit Windows 10 ($N=5000$), now takes a fraction of a second.\\

\end{small}

\section{Introduction}
\label{introduction}

A broadly-applicable rank-based approach for detection and extraction of generally non-linear trends in noisy time series has recently been introduced \cite{ierley2019universal} and we shall now briefly review the mathematical essentials. The input time series is segmented into $n_{t}$ samples, with each sample having $n_{T}$ data points. A square $n_{T}\times n_{T}$ population matrix $P$ is then calculated such that $P_{j,k}$ is the population (number) of data points with order $j$ (position in the sample), and rank $k$ (position in the sample after an ascending sort)\cite{ierley2019universal}. Alternatively, $P_{j,k}$ can also be viewed as a 2D probability density function (pdf) or a histogram over the plane defined by rank and times axes. The matrix $P$ is illustrated below in \eqref{SketchofP}.

\begin{align}\label{SketchofP}
\begin{split}
P &= \left( \begin{array}{ccccc|ccc}
    P_{1,1} & P_{1,2} & \dots & P_{1,k-1} & P_{1,k} & P_{1,k+1} & \dots & P_{1,n_{T}} \\
    P_{2,1} & P_{2,2} & \dots & P_{2,k-1} & P_{2,k} & P_{2,k+1} & \dots & P_{2,n_{T}} \\
    \vdots & \vdots &  & \vdots & \vdots & \vdots &  & \vdots\\
    P_{j-1,1} & P_{j-1,2} & \dots & P_{j-1,k-1} & P_{j-1,k} & P_{j-1,k+1} & \dots & 
    P_{j,n_{T}} \\
    P_{j,1} & P_{j,2} & \dots & P_{j,k-1} & P_{j,k} & P_{j,k+1} & \dots & P_{j,n_{T}} \\
    \hline
    P_{j+1,1} & P_{j+1,2} & \dots & P_{j+1,k-1} & P_{j+1,k} & P_{j+1,k+1} & \dots & P_{k+1,n_{T}} \\
    \vdots & \vdots &  & \vdots & \vdots & \vdots &  & \vdots\\
    P_{n_{T},1} & P_{n_{T},2} & \dots & P_{n_{T},k-1} & P_{n_{T},k} & P_{n_{T},k+1} & \dots & P_{n_{T},n_{T}} \\
    \end{array}
    \right)\\
    \end{split}
\end{align}

To ``zoom in'' on the trends hidden in $P$, the $Q$-transform was introduced  \cite{ierley2019universal} as follows

\begin{equation}\label{QofPequation1}
\begin{split}
    Q_{j,k}&=\Bigg(\frac{\sum_{m=1}^{j}\sum_{n=1}^{k}P_{m,n}+\sum_{m=j+1}^{n_{T}}\sum_{n=k+1}^{n_{T}}P_{m,n}}{jk+(n_{T}-j)(n_{T}-k)} \\
    &\qquad\qquad-\frac{\sum_{m=1}^{j}\sum_{n=k+1}^{n_{T}}P_{m,n}+\sum_{m=j+1}^{n_{T}}\sum_{n=1}^{k}P_{m,n}}{(n_{T}-j)k+j(n_{T}-k)}\Bigg)\frac{n_{T}}{n_{t}}
    \end{split}
\end{equation}

To understand the construction, consider the division of $P$ into quadrants for calculation of $Q_{j,k}$ as shown on the RHS of \eqref{SketchofP}.
Each element of $Q$ is the difference between the average matrix element of the combined upper left and lower right quadrants of $P$, and the average matrix element of the combined upper right and lower left quadrants, normalized by the overall average matrix element of $P$, $\left<P\right>\equiv\sum_{m=1}^{n_{T}}\sum_{n=1}^{n_{T}}P_{m,n}/n_{T}^{2}=n_{t}/n_{T}$. 

The number of operations ($+,-,\times,\div$) required to compute $Q$ using equation \eqref{QofPequation1} is ${\cal O}(n_{T}^{4})$. For large $n_{T}$ and repeated calls, as will often be needed in applications, the computation time can become prohibitively long. In addition, setting $\left<Q\right> = 0$ where angular brackets denote average over matrix elements, functions as a trend detector when the functional form of the trend is not available and we shall illustrate it on the time series of cosmic rays in \ref{cosmicrayillustration}.  To that end, our purpose in this paper is four-fold: 

(i) present a ${\cal O}(n_{T}^{2})$ algorithm for computing the $Q$-transform and its MATLAB implementation;

(ii) supply an open source (Julia) implementation;

(iii) present an efficient calculation of $\left<Q\right>$, where $\left<Q\right>$ is the average matrix element of $Q$. The departure of this (scalar) quantity from zero is used to detect presence of trend\cite{ierley2019universal}.

(iv) provide an illustrative example from a long cosmic ray time series; 

\section{Derivation of the Algorithm}\label{derivationalgorithmsection}

To begin, note that equation \eqref{QofPequation1} can be simplified by making use of constraints on $P$ that each row and column sum to $n_{t}$. Thus, the sums of elements in the four quadrants of P, entering the numerator of \eqref{QofPequation1} are not independent. Numbering the quadrants as 1-4 beginning from the upper right, moving counter-clockwise, and calling the sums of elements in each quadrant $i$ as $\Sigma^{(i)}$, we have

\begin{equation}
\begin{split}
    \Sigma^{(1)}+\Sigma^{(2)}&=jn_{t} \\
    \Sigma^{(2)}+\Sigma^{(3)}&=kn_{t} \\ 
    \Sigma^{(3)}+\Sigma^{(4)}&=(n_{T}-j)n_{t} \\
    \Sigma^{(4)}+\Sigma^{(1)}&=(n_{T}-k)n_{t}
\end{split}
\end{equation}

This system of four equations in four unknowns ($\Sigma^{(i)}$, $i=1-4$) is under-determined and when recast as a 4x4 matrix equation, has a matrix rank of three. Thus, only one of the four $\Sigma^{(i)}$ is independent and we picked $\Sigma^{(2)}$ for that purpose.

\begin{equation}
\begin{split}
    \Sigma^{(1)}&=jn_{t}-\Sigma^{(2)} \\
    \Sigma^{(2)}&=\Sigma^{(2)} \\
    \Sigma^{(3)}&=kn_{t}-\Sigma^{(2)} \\
    \Sigma^{(4)}&=(n_{T}-j-k)n_{t}+\Sigma^{(2)}
    \end{split}
\end{equation}

This can be substituted back into equation \eqref{QofPequation1}, making explicit a $(j,k)$ index on $\Sigma^{(i)}$.


\begin{equation}
\begin{split}
    Q_{j,k}&=\Bigg(\frac{(n_{T}-j-k)n_{t}+2\Sigma^{(2)}_{j,k}}{jk+(n_{T}-j)(n_{T}-k)}-\frac{(j+k)n_{t}-2\Sigma^{(2)}_{j,k}}{(n_{T}-j)k+j(n_{T}-k)}\Bigg)\frac{n_{T}}{n_{t}}
\end{split}
\end{equation}

Define $D$ as a $(n_{T}-1)\times(n_{T}-1)$ matrix, whose elements are the product of the two denominators in equation \eqref{QofPequation1}:

\begin{equation}\label{DEquation}
    D_{j,k}=(jk+(n_{T}-j)(n_{T}-k))(j(n_{T}-k)+(n_{T}-j)k)
\end{equation}

The $Q$ matrix can be expressed compactly in terms of $\Sigma^{(2)}_{j,k}$ and $D_{j,k}$.

\begin{equation}\label{QDofSigma2}
    \begin{split}
        Q_{j,k}D_{j,k}&=\frac{2n_{T}^{3}}{n_{t}}\Bigg(\Sigma^{(2)}_{j,k}-\frac{n_{t}}{n_{T}}jk\Bigg)
    \end{split}
\end{equation}

The motivation for this is that the second quadrant sum $\Sigma^{(2)}_{j,k}$ satisfies a recurrence relation.

\begin{equation}
    \Sigma^{(2)}_{j,k}=\Sigma^{(2)}_{j-1,k}+\Sigma^{(2)}_{j,k-1}-\Sigma^{(2)}_{j-1,k-1}+P_{j,k}
\end{equation}

Taken together with equation \eqref{QDofSigma2}, this yields a recurrence relation for $Q$.

\begin{equation}\label{QRecursion}
    \begin{split}
        Q_{j,k}&=\frac{1}{D_{j,k}}\Bigg(D_{j,k-1}Q_{j,k-1}+D_{j-1,k}Q_{j-1,k} \\
        &\qquad\qquad-D_{j-1,k-1}Q_{j-1,k-1}+\frac{2n_{T}^{3}}{n_{t}}(P_{j,k}-\frac{n_{t}}{n_{T}})\Bigg)
    \end{split}
\end{equation}

The algorithm used to calculate $Q$ via \eqref{QRecursion} is described in Algorithm \ref{algorithmforQ} and its MATLAB and Julia implementations accompany this manuscript. The full $Q$ matrix is calculable in ${\cal O}(n_{T}^{2})$ operations, as is seen by observing that each element of $Q$ can be calculated in ${\cal O}(1)$ from a small number of neighboring $Q$ elements and some constants, and that the total number of elements in $Q$ is $(n_{T}-1)^2$. \\

\begin{algorithm}[H]
\KwData{$n_{T}\times n_{T}$ population matrix $P$} 
\KwResult{$(n_{T}-1)\times(n_{T}-1)$ matrix $Q$}
First, verify that input matrix $P$ satisfies row and column sum constraints\;
Calculate $(1,1)$ element using equation \eqref{QofPequation1}\;
Evaluate the remainder of the first row of $Q$ using recursion. Zero or negative indices of $Q$ in equation \eqref{QRecursion} mean $Q$ may be replaced by zero there, as a more careful consideration of values of $Q$ near the matrix edge confirms\;
Repeat for the remainder of the first column of $Q$\;
Calculate (2,2) from surrounding three elements\;
Evaluate remainder of row/column 2 of $Q$ recursively\;
\vdots
Calculate $(n,n)$ element of $Q$ from upper left surrounding elements\;
Evaluate remainder of row/column $n$ of $Q$, using recursion\;
\vdots
Calculate $(n_{T}-1,n_{T}-1)$ element
\caption{An ${\cal O}(n_{T}^{2})$ implementation of the $Q$-transform, using recurrence. The (1,1) element is found first. Rows and columns are found by moving rightwards or downwards from diagonal elements, which are found from previous rows/columns.}
\end{algorithm}\label{algorithmforQ} 

\section{Analytical Results}
\label{analyticalresults}

One key result of this paper is equation \eqref{QRecursion}, just derived. This permits an ${\cal O}(n_{T}^{2})$ method for calculating $Q$ that is much faster than the ${\cal O}(n_{T}^{4})$ brute force evaluation of equation \eqref{QofPequation1}, especially for large $n_{T}$. Another essential result is the transformation for $P$, given $Q$.  This was obtained by rearranging equation \eqref{QRecursion} as follows. 

\begin{equation}\label{PofQ}
    \begin{split}
        P_{j,k}=\Bigg(D_{j,k}Q_{j,k}-D_{j,k-1}Q_{j,k-1}-D_{j-1,k}Q_{j-1,k} \\
        +D_{j-1,k-1}Q_{j-1,k-1}\Bigg)\frac{n_{t}}{2n_{T}^{3}}+\frac{n_{t}}{n_{T}}
    \end{split}
\end{equation}

Not only does this transformation turn out to be stably computable but also efficiently so. In fact, it can be accomplished also in $O(n_{T}^{2})$ operations and is implemented in MATLAB and Julia programs in the accompanying files. These results allow analyses of previously inaccessible data because of the prohibitively long computation times.  The confirmation of the speed up of equation \eqref{QRecursion} over equation \eqref{QofPequation1} directly in terms of CPU time is given in Fig. \ref{fig:bruteforcevsrecurrence} below. 

\begin{figure}[ht!]
    \centering
    \includegraphics[width=0.5\textwidth]{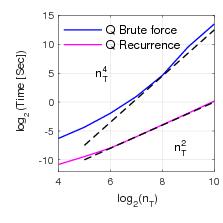}
    \caption{A comparison of calculation times for the Q-transform. A recursive approach reduces the CPU time from ${\cal O}(n_{T}^{4})$ to ${\cal O}(n_{T}^{2})$. The blue curve shows results from an implementation of the brute force approach of equation \eqref{QofPequation1}, while the red curve shows timing data from the optimized calculation of equation \eqref{QRecursion}. The contrast in complexity order becomes apparent for larger size matrices, $\log_{2}(n_{T})\gtrapprox4$. Note that the calculation time becomes intractable for $n_{T}\gtrapprox2^{10}\approx1024$ for the brute force calculation\cite{laptopdetails}.}
    \label{fig:bruteforcevsrecurrence}
\end{figure}

As a sample application, possible because of the computational improvement provided by calculating the $Q$-transform recursively rather than by the direct evaluation of the double sums in equation \eqref{QofPequation1}, we choose $n_{T}=5000\approx2^{12.3}$, which lies just outside of the axis range shown in Fig. \ref{fig:bruteforcevsrecurrence}. A calculation using the recursive result in equation \eqref{QRecursion} takes about 0.7 seconds\cite{laptopdetails}. In comparison, using Fig. \ref{fig:bruteforcevsrecurrence} to extrapolate the $O(n_{T}^{4})$ curve out to $\log_{2}(n_{T})=12.3$, a brute force calculation would take approximately 80 days and hence is not shown in the figure.

\section{Fast Algorithm for Calculating $\left<Q\right>$}\label{fastQavgsection}

We now turn to efficient calculation of $\left<Q\right>$, the mean matrix element of $Q$ in the special case of large $n_{T}$ and small $n_{t}$.  In such single sample (time series) cases, the matrix $P$ is sparse, consisting of ${N_T}^2 - N_T$ zeroes and need not be stored in memory in its entirety.  Rather, only indices of the non-zero matrix elements, found by independently sorting each of the repeated $n_{t}$ trials, may suffice to calculate $Q$. Also, it was shown in \cite{ierley2019universal} that the entire $Q$ information is not required when one is concerned merely with trend detection or parameter estimation of a linear trend. One such application (time series of cosmic ray arrivals) is discussed in the next section. In such cases, it suffices to calculate only element-averaged metric

\begin{equation}\label{Qavg}
    \left<Q\right> \equiv \frac{1}{(n_{T}-1)^{2}}\sum_{j,k=1}^{n_{T}-1}Q_{j,k}
\end{equation}

A sufficiently large value of $\left<Q\right>$ implies the presence of a trend. Here, ''sufficiently large'' is with reference to a fiducial value expected for pure noise, a formula for which in terms of $n_{T}$ and $n_{t}$ is given in \cite{ierley2019universal}. Using the results of Section \ref{analyticalresults}, $Q$, and thus $\left<Q\right>$ (which has $(n_{T}-1)^{2}$ terms), can be computed from $P$ in ${\cal O}(n_{T}^{2})$ operations.  In order to calculate $\left<Q\right>$, only accumulated mean value is needed and the rest of the matrix elements need not be stored, thereby reducing  complexity of the calculation. To that end, it is shown in \cite{ierley2019universal} that $\left<Q\right>$ can be written as the sum of elements of the Hadamard product of a $(n_{T}\times n_{T})$ matrix $m$ and $P$ itself (Appendix B.3 of \cite{ierley2019universal}). The matrix $m$ is not determined by the data, and depends solely on $n_{t}$ and $n_{T}$, the former being only an inverse multiplicative constant. Once $m$ is constructed, the mean element of Q is easily accessed as $\left<Q\right>=\sum_{j=1}^{n_{T}} \sum_{k=1}^{n_{T}} m_{j,k}P_{j,k}$. For sparse $P$, $n_{t}\ll n{T}$, this means a calculation of order $n_{T}$, much faster than the ${\cal O}(n_{T}^{2})$ needed to calculate the matrix $Q$ explicitly prior to averaging. As it stands, $m$ takes ${\cal O}(n_{T}^{4})$ to construct, which can become prohibitive for large $n_{T}$. Not only this, but $m$ is memory limited to about $n_{T}=\sqrt{10}\times10^4$ for an 8GB RAM, being of type double (8 bytes per element). Since for $n_{t}\ll n_{T}$ the matrix $P$ is sparse, in this case we only need calculate the elements of $m$ corresponding to nonzeros in $P$. This greatly reduces demands on memory, allowing $n_{T}$, the number of data points per trial, to be as large as about $10^9$ for $n_{t}=1$ and a 8 GB RAM. Also, there is an ${\cal O}(n_{T})$ way to approximate any element of $m$ in ${\cal O}(1)$, operations, reducing the calculation of $m$ for sparse $P$ to ${\cal O}(n_{T})$. For nonsparse $P$, when the full matrix $m$ is needed, the approximation scheme gives $m$ in ${\cal O}(n_{T}^{2})$, due the number of elements needed. We also find that $m$ may be calculated exactly, up to accumulated rounding errors due to finite machine precision, by an ${\cal O}(n_{T}^{2})$ recursive algorithm, much like for $Q$ in section \ref{derivationalgorithmsection}. The advantage in this case is that $m$ need only be calculated once, for each $n_{T}$, and then it applies to any dataset of the same size parameter $n_{T}$, modulo a rescaling due to varying $n_{t}$. This saves the time needed for calculating $Q$ itself each time. The disadvantage of using recursion to find $m$ as compared with using the approximate approach is that recursion can only create contiguous rectangular blocks of $m$. In contrast, the approximation for $m$ allows only the needed elements to be calculated, regardless of how they are spatially related in the matrix. To this we now turn.



To describe how the matrix $m$ is approximated, which is the most efficient way to calculate $\left<Q\right>$ for large $n_{T}$ and $n_{t}\ll n_{T}$, we first note the following: the rank one matrix that is an outer product of the first column of $m$ with itself, normalized by $1/m(1,1)$, provides a fair approximation of the entire matrix $m$. Seeing that this approximation is rank one suggests the approximation can be improved by adding another rank one term. This turns out to be so. One simply takes a linear combination of the first two columns of $m$, and since the first row and column of $m$ are already exact, chooses this combination such that a new vector is obtained with vanishing first element. The outer product of this vector with itself is zero along the first row and column, and thus does not alter the previous exact rank one approximation there. If one then adds this new rank one matrix to the old, while choosing a scalar coefficient to match any of the elements in the original second column of $m$, one obtains a rank two approximation of $m$ that is exact in the first \textit{two} rows and columns. This holds true for any symmetric matrix, as a little algebra can show. Moreover, this extends easily: a third ''basis vector'' may be obtained with vanishing first and second elements by judiciously mixing the first three exact columns of $m$. Again adding the outer product of this new vector with itself to the rank 2 approximation, with an appropriate constant chosen to match an arbitrary element of the exact third column of $m$, a rank 3 approximation is obtained that is exact in the first three rows and columns. And so on. By generalization, beginning with $r$ columns of $m$ yields a rank $r$ approximation, exact in the first $r$ rows and columns of $m$. Since $m$ always has the same form when regarded as a two dimensional function, the effect of increasing the number of rows/columns $n_{T}$ is to bring nearby columns closer numerically. Therefore, practically, difficulties arise with the above approximation scheme due to nearby columns of $m$ becoming linearly dependent for large $n_{T}$, but these can be circumvented by avoiding successive columns, but picking columns increasingly separated with $n_{T}$, so as to roughly maintain proportionate horizontal locations in the matrix $m$. It also proves advantageous to mix the columns such that the zeros in the new column vectors are also spaced out proportionately within the matrix and not merely adjacent and at the beginning. Heuristically, the optimal case is when the zeros in the new column vectors have the same spacing as the columns. This improves the conditioning of a certain matrix that must be inverted in this process. Also, we find that for uniform column spacing, there is an optimal rank approximation of $r=6$. In this case, the optimal column/row zero spacings $s$ are given empirically by $s=[0.92143543+0.02465247\times n_{T}]$. Taking non-uniformly spaced columns of the matrix $m$ yields generally much better results, as found for example by using MATLAB's Genetic Algorithm to find the optimal columns of $m$ for a rank $r=6$ expansion with row zero spacings matching the column spacings. We also choose the coefficients of the rank 1 matrices in order to optimize the approximation of $m$ along its diagonal, via least squares (MATLAB's backslash operator).

We note that this approximation is essentially a low rank matrix approximation that uses low rank matrix completion, topics that both arise in data science\cite{nguyen2019low}.

With this approximation of $m$, we now show that the cost of computing $m$ is reduced from ${\cal O}(n_{T}^{2})$ to ${\cal O}(n_{T})$ for the overhead, and ${\cal O}(1)$ operations for each element after that. We also show that the memory overhead is also ${\cal O}(n_{T})$, plus the cost of each element computed after that (between ${\cal O}(n_{T})$ and ${\cal O}(n_{T}^{2})$).

\begin{figure}[ht!]
    \centering
    \includegraphics[width=0.5\textwidth]{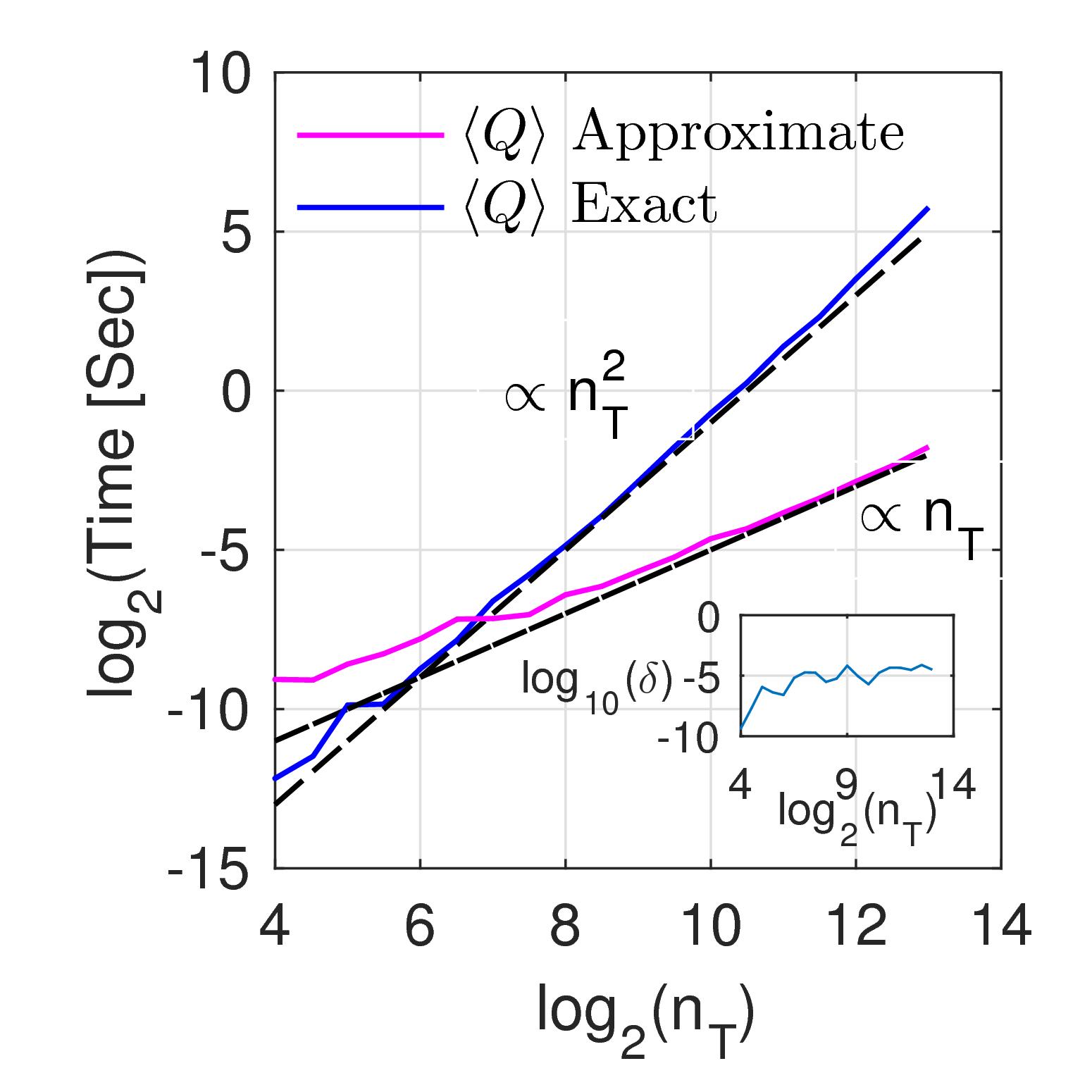}
    \caption{${\cal O}(n_{T})$ vs ${\cal O}(n_{T}^{2})$ calculations of $\left<Q\right>\equiv$ average over all elements of $Q$, in the limit $n_{t}<<n_{T}$. The relative difference $\delta \equiv (<Q>_{approx.} - <Q>_{exact})/<Q>_{exact}$ is shown in the inset. The linear time includes the time to "precompute" the needed subset of elements of the matrix $m$ (see the text) and take the Hadamard product with $P$. The quadratic time includes the time to calculate, store, and average the full Q matrix.}
    \label{fig:Qavgexactvsapprox}
\end{figure}

The equation for $m$ is now derived. As shown in \cite{ierley2019universal}, unwrapping matrices $P$ and $Q$ to vectors allows to write $Q=MP$, where $M$ is a matrix with dimensions $(n_{T}-1)^2\times n_{T}^2$. $m$ is then defined by $m=1^{T}M=[1 1 1 ... 1]M$ and has dimension $1\times n_{T}^2$. In this paper, by comparing with equation \eqref{QofPequation1} and carefully converting between matrix indices and linear vector indices, we find a closed form solution for $m$ rewrapped as a $n_{T}\times n_{T}$ matrix, and satisfying $\left<Q\right>=\sum_{j,k}(m \circ P)_{j,k}$, where $\circ$ denotes the Hadamard product.

\begin{equation}\label{mjk}
    \begin{split}
        m_{j,k}=\frac{n_{T}}{n_{t}(n_{T}-1)^2}\bigg(&\sum_{m=1}^{j-1}\sum_{n=1}^{k-1}d_{m,n}^{(1)}+\sum_{m=j}^{n_{T}-1}\sum_{n=k}^{n_{T}-1}d_{m,n}^{(1)}-\\
        &\quad \sum_{m=1}^{j-1}\sum_{n=k}^{n_{T}-1}d_{m,n}^{(2)}-\sum_{m=j}^{n_{T}-1}\sum_{n=1}^{k-1}d_{m,n}^{(2)}\bigg)
    \end{split}
\end{equation}

\begin{equation}\label{dmnequations}
\begin{split}
    d_{m,n}^{(1)}&=\frac{1}{mn+(n_{T}-m)(n_{T}-n)}\\
    d_{m,n}^{(2)}&=\frac{1}{m(n_{T}-n)+(n_{T}-m)n}
\end{split}
\end{equation}

Since only the non-zero entries of $P$ contribute to the element-wise product with $m$, and $P$ can have as few as ${\cal O}(n_{T})$ non-zero entries (when $n_{t}=1$), $\left<Q\right>$ can be computed in as low as ${\cal O}(n_{T})$ operations, after overhead that is also ${\cal O}(n_{T})$, leaving a grand total of ${\cal O}(n_{T})$ operations. This is in contrast to the ${\cal O}(n_{T}^{2})$ operations it would take to compute $Q$ using the recursive algorithm of section \ref{derivationalgorithmsection}, and then sum and average the elements of $Q$. This difference is illustrated in figure \ref{fig:Qavgexactvsapprox}.

\section{Illustration on Time Series of Cosmic Rays}\label{cosmicrayillustration}

\begin{figure}[ht!]
    \centering
    \includegraphics[width=0.5\textwidth]{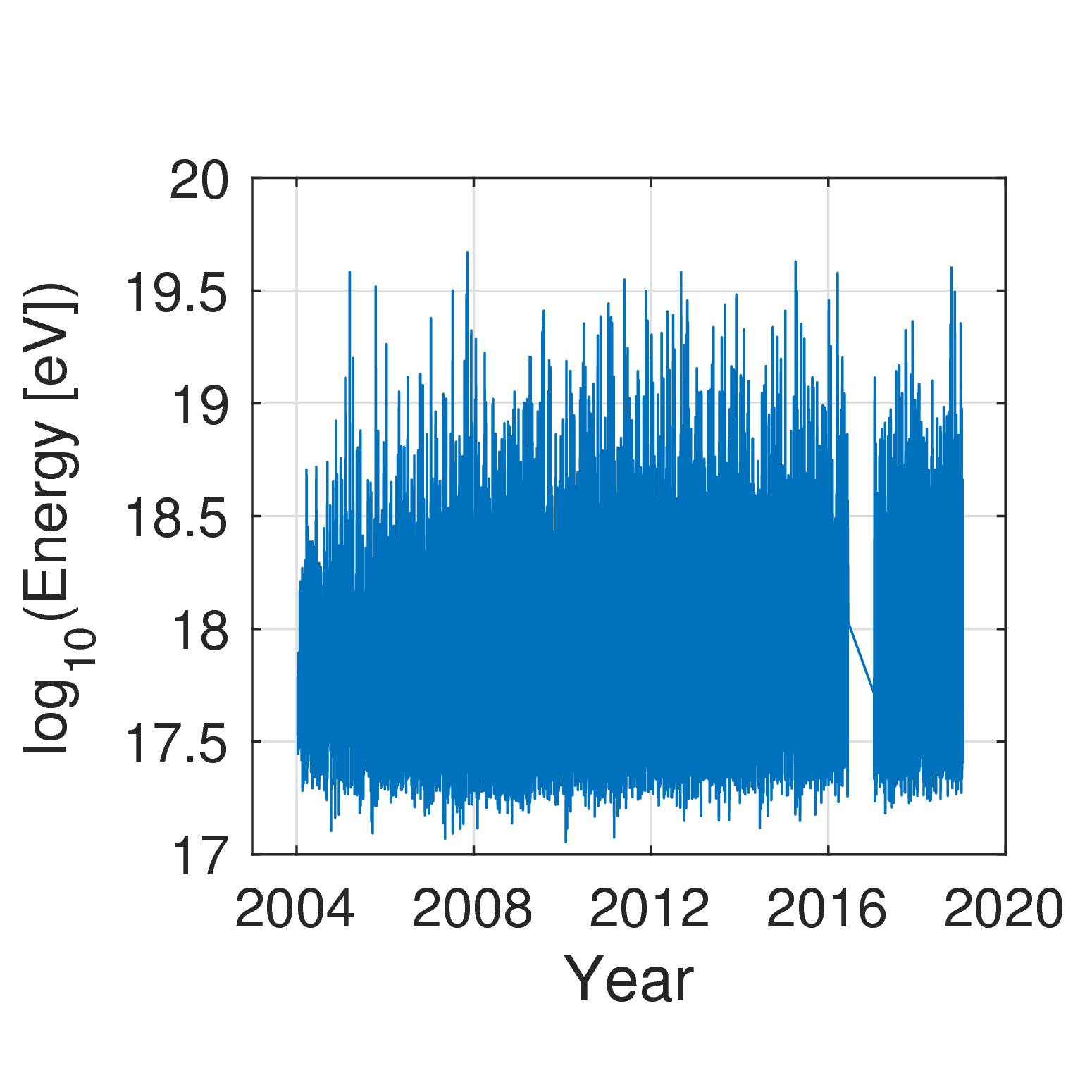}
    \caption{Time series of energies of high energy cosmic rays. There are a total of 49223 events between 2004 and 2019. Energy scale is $10^{18}\text{eV}=1\text{EeV}$. The data is uncorrelated (white), and has a non Gaussian distribution. Data is taken from the Pierre Auger Observatory Public Event Explorer \cite{AugerPublicData2019}, and represents 1\% of all data taken by the observatory.}
    \label{fig:CosmicRayTimeSeries}
\end{figure}

To illustrate the importance of the numerical acceleration for trend detection as just described in the previous section, we pick an example from cosmic ray physics.  The data consists of 49,223 events (only 1\% of the total data is available to general public), in a form of a time series of arrivals with various energies (see Fig. \ref{fig:CosmicRayTimeSeries} from data in \cite{AugerPublicData2019}).

Energy-resolved flux (spectrum) plays the central role in the field and it is universally assumed that the underlying time series are statistically stationary.  Are they?  Here we ask whether the time series in Fig.\ref{fig:CosmicRayTimeSeries} are stationary and we use $\left<Q\right>$ to test the hypothesis. Stationarity implies that $\left<Q\right> \approx 0$ and the significance of the deviation is judged in units of the standard deviation of steady value $\left<Q\right>=0$ via the asymptotics in equation (10) of reference \cite{ierley2019universal}.  Calculation of the auto-correlation function for this cosmic ray data shows that it is uncorrelated (``white'') so using $\left<Q\right>$-asymptotics is particularly relevant.

For sake of consistency, we tested a variety of data partitioning, but with the same product $n_{t}n_{T}$. Table \ref{tab:cosmictabletimings} shows the importance of the approximate $\left<Q\right>$ algorithm. For $n_t$ approaching unity, dimensions of $Q$ are $\sim 10^4 \times 10^4$ and the $O(N)$ algorithm is crucial. Table \ref{tab:cosmictablevalues} shows the calculations. To our surprise, the $\left<Q\right>$-test consistently detects a presence of a trend beyond reasonable doubt. Specifically, $\left<Q\right>$ = $0.06$ gives the confidence limit of 19$\sigma$ (taking the case $n_t\gtrapprox100$ for specificity).  The associated linear trend is large enough to affect the spectrum and cast doubt on the traditional power-law analysis as the latter implies stationarity via the Wiener-Khintchin theorem.
\begin{table}[ht!]
\centering
\resizebox{\textwidth}{!}{%
\begin{tabular}{@{}ccccc@{}}
\toprule
$n_{T}$ & Time $Q$ (Sec) & Time $Q$ sum (Sec) & \begin{tabular}[c]{@{}c@{}}Time m Basis and\\Coefficients (Sec)\end{tabular} & Time $m_{i,j}P_{i,j}$ sum (Sec) \\ \midrule
49140 & - & - & 0.968 & 1.58E-02 \\
24570 & - & - & 0.484 & 1.15E-02 \\
9828 & 31.8 & 8.70E-02 & 0.200 & 1.14E-02 \\
4095 & 2.85 & 1.53E-02 & 8.39E-02 & 7.37E-03 \\
468 & 1.91E-02 & 3.95E-04 & 1.05E-02 & 7.02E-03 \\
91 & 5.50E-04 & 2.94E-05 & 2.73E-03 & 7.69E-03 \\
12 & 4.99E-05 & 2.34E-05 & 1.22E-03 & 7.11E-03 \\ \bottomrule
\end{tabular}%
}
\caption{\label{tab:cosmictabletimings}To investigate the range of possibilities, we compare several partitions of the time-series of cosmic rays. By trimming the data from 49223 down to 49140 datapoints, the number of distinct integers $n_{t}$, $n_{T}$ such that $n_{t}n_{T}$ equals the total number of kept data points is maximized, providing many partitions for study. The product $n_{t}n_{T}$ is held constant. In this table, timing dependence of partitions is shown. For all entries shown, $n_{t}=49140/n_{T}$ is integer. One can see that the fast approximate calculation for $\left<Q\right>$ is possible for all possible partitions, while those based on full calculation of $Q$ become memory limited past about $n_{T}=10^4$, as indicated by the dashes. The approximate calculation stores neither $P$ nor $Q$, and is able to be performed up to the maximum $n_{T}$. Thus, the approximate method is the only way to probe these partitions. For $n_{T}$ greater than a few hundred, the approximate calculation, consisting of approximating basis vectors and coefficients for $m$, followed by a Hadamard product with $P$, is much faster than direct evaluation of $Q$ and its mean element.}
\end{table}

\begin{table}[ht!]
\centering
\resizebox{\textwidth}{!}{%
\begin{tabular}{@{}ccccc@{}}
\toprule
$n_{T}$ & $\left<Q\right>_{exact}$ & $\left<Q\right>_{approximate}$ & $\delta$ & $\left<Q\right>/\sigma_{white\;noise}$\\ \midrule
49140 & - & 6.09E-02 & -  & 19.0\\
24570 & - & 6.09E-02 & - & 19.0 \\
9828 & 6.10E-02 & 6.10E-02 & 3.65E-06 & 19.0 \\
4095 & 6.09E-02 & 6.09E-02 & 2.13E-06 & 19.0 \\
468 & 6.12E-02 & 6.12E-02 & -2.92E-06 & 19.1 \\
91 & 6.22E-02 & 6.22E-02 & -7.95E-07 & 19.1 \\
12 & 7.27E-02 & 7.27E-02 & -1.78E-15 & 18.7 \\ \bottomrule
\end{tabular}%
}
\caption{Companion to Table \ref{tab:cosmictabletimings} for $\left<Q\right>$ calculation values under different data partitions. Both exact and approximate calculations give results that are roughly independent of data partition for sufficiently large $n_{T}\gtrapprox10^2$. For all entries, $n_{t}=49140/n_{T}$. Accuracy of the approximate calculation is $\delta=\left<Q\right>_{approximate}/\left<Q\right>_{exact}-1$, and is excellent, being better than $10^{-6}$ where $n_{T}$ is small enough that the exact calculation can be performed and compared to. $\left<Q\right>$, normalized by its standard deviation from zero for a comparable white noise process, is about $18$, indicating the presence of a trend in the data.}
\label{tab:cosmictablevalues}
\end{table}

\section{Concluding Remarks}
\label{concluding remarks}
In conclusion, we have discovered an $O(N^{2})$ calculation of a previously $O(N^{4})$ matrix transform with applications in trend detection from noisy data. This increases the efficiency of the transform, and allows access to previously out-of-reach data sample lengths $N$. For the special case of a small number of samples $n_{t}$, we present also an $O(N)$ calculation of trend detection metric $\left<Q\right>$ which bypasses the need to carry out the full $Q$-transform. Open access computer codes are provided for both of these calculations.

\section{Declaration of Interests}
The authors have no competing interests to declare.

\section{Funding}
This work was supported by the National Science Foundation grant AGS-1639868.
\appendix

\section{Mathematical Identities for matrix m used in Software}

From \eqref{mjk}, it can be shown necessarily that $m_{j,k}=m_{k,j}$, a symmetric matrix, and $m_{j,n{T}-k+1}=-m_{j,k}$, a matrix odd under vertical or horizontal inversion. For the upper left matrix quadrant $j\leq n_{T}-j$ and $k\leq n_{T}-k$ it can be shown from \eqref{mjk} that the following is necessary:

\begin{multline}\label{mjksimplified}
    m_{j,k}=\frac{n_{T}}{n_{t}(n_{T}-1)^2}\bigg(2\sum_{m=\lfloor\frac{n_{T}}{2}\rfloor+1}^{n_{T}-j}\frac{\psi_{0}(k-\frac{n_{T}(n_{T}-m)}{n_{T}-2m})-\psi_{0}(1-k-\frac{mn_{T}}{n_{T}-2m})}{n_{T}-2m}\\
    +\frac{2}{n_{T}}(n_{T}-2k+1)\left|cos\left(\frac{n_{T}\pi}{2}\right)\right|\bigg)
\end{multline}

Here, $\psi_{0}$ is the polygamma function of order 0 (e.g. MATLAB psi function, Julia module SpecialFunctions' polygamma function with zero as the first argument). From this, the following can be shown:

\begin{equation}\label{mjkrowrecurrence}
    m_{j+1,k}=m_{j,k}-\frac{2n_{T}}{n_{t}(n_{T}-1)^2}\frac{\psi_{0}(k-\frac{n_{T}(n_{T}-j)}{n_{T}-2j})-\psi_{0}(1-k-\frac{jn_{T}}{n_{T}-2j})}{n_{T}-2j}
\end{equation}

This allows recursion down the columns of m in an exact calculation. Let the subtracted quantity from $m_{j,k}$ be $f(j,k)$. The following is again necessary:

\begin{equation}\label{mjkdiagrecurrence}
    m_{j,j}=m_{j+1,j+1}+f(j,j)+f(j,j+1)
\end{equation}

This permits recurrence along the diagonal of m in an exact calculation. Finally, it also is necessary that $m$ satisfy the following:

\begin{equation}\label{mjkfullrecurrence}
    m_{j,k}+m_{j-1,k-1}-m_{j,k-1}-m_{j-1,k}=2(d_{j-1,k-1}^{(1)}+d_{j-1,k-1}^{(2)})
\end{equation}

Here $d_{j,k}^{(1)}$ and $d_{j,k}^{(2)}$ are given in equation \eqref{dmnequations} of the main text. This causes $m$ as a whole to be calculable in ${\cal O}(n_{T}^{2})$ operations, in contrast to ${\cal O}(n_{T}^{4})$ from equation \eqref{mjk} (${\cal O}(n_{T}^{2})$ per element).



\bibliographystyle{elsarticle-num}
\bibliography{Manuscript_AppendixReplaced}

\begin{thebibliography}{1}
\bibitem{laptopdetails}
All calculations were performed with the first author's personal Dell Inspiron
  15 laptop, equipped with 2.2GHz AMD quadcore processor, using 64 bit Windows
  10. The best time scalings were provided by the slower 32 bit version of
  MATLAB, and thus this is what is shown. For Tables
  \ref{tab:cosmictabletimings} and \ref{tab:cosmictablevalues}, 64 bit MATLAB
  was used, being the faster version.
\end{thebibliography}







\end{document}